\documentclass[aps,nofootinbib,11pt,preprintnumbers]{revtex4}
\usepackage{amssymb,amsmath}
\usepackage{graphicx}
\newcommand{\bea}{\begin{eqnarray}}
\newcommand{\eea}{\end{eqnarray}}

\def\x{{\bf x}}
\def\E{{\bf E}}
\def\K{{\bf K}}
\def\F{{\bf F}}
\def\0{{\bf 0}}

\def\p{{\bf p}}

\begin{document}
\preprint{MIT-CTP 4049}

\title{
Inhomogeneous phases in the Nambu-Jona--Lasino and quark-meson model
}
\author{Dominik Nickel}
\affiliation{Center for Theoretical Physics,
  Massachusetts Institute of Technology,
  Cambridge, MA 02139, USA}
\date{June 2009}

\begin{abstract}
\noindent
We discuss inhomogeneous ground states of the Nambu-Jona--Lasino (NJL) and quark-meson (QM) model within mean-field approximation and their possible existence in the respective phase diagrams.
For this purpose we focus on lower-dimensional modulations and point out that known solutions in the 2+1 and 1+1 dimensional (chiral) Gross-Neveu (GN) model can be lifted to the to the 3+1 dimensional NJL model.
This is worked out in detail for one-dimensional modulations and numerical results for the phase diagrams are presented.
Focus is put on the critical point and on vanishing temperatures.
As an interesting result the first order transition line in the phase diagram of homogeneous phases gets replaced by an inhomogeneous phase which is bordered by two second order transition lines.
\end{abstract}

\maketitle

\section{introduction}
\label{sec:intro}
\noindent
Until today the phase diagram of quantum chromodynamics (QCD) is subject to intense theoretical and experimental investigations (for dedicated reviews see Ref.~\cite{reviews}).
Experimentally its structure at finite temperatures is explored by heavy ion collisions, which currently focus on the formation and properties of a strongly interacting plasma at large temperatures as well as on the search for a chiral critical point in the phase diagram.
Theoretically ab initio calculations are limited to small net-baryon densities and as a consequence possible scenarios at non-vansihing densities and strong coupling are often discussed within phenomenological models.

\noindent
Since NJL-type models\footnote{We refer to NJL-type models as models that at least in the applied approximation reduce to the NJL model, possibly extended by additional point-like interactions,  on a technical level. This includes e.g. simplified ans\"atze for the gluon interaction, the use of different regularizations, the instanton liquid model and the QM model.} share global symmetries and the phenomenon of chiral symmetry breaking with QCD, they are - typically in mean-field approximation - widely used to study the phase diagram at moderate temperatures and densities.
In this context e.g. color-superconductivity~\cite{reviews,Alford:2007xm} as well as the location of the critical point(s) have been addressed~\cite{Stephanov:2007fk}. 
In recent years this branch of models has been extended to Polyakov-NJL-type models, which include the dynamics of the Polyakov-loop as the order parameter of confinement~\cite{Fukushima:2003fw, Ratti:2005jh}. However for vanishing temperatures these models reduce to the respective NJL-type model.

\noindent
In this work we investigate inhomogeneous ground states in the NJL~\cite{Nambu:1961tp} and QM model~\cite{GellMann:1960np,Scavenius:2000qd}. These are characterized by a spatially varying order parameter and have been discussed for QCD at least in the large $N$ limit, where they are expected to form the ground state at sufficiently high densities~\cite{Deryagin:1992rw,Shuster:1999tn,Park:1999bz}. Related to this they show up in holographic models~\cite{Rozali:2007rx} and in the quarkyonic matter picture~\cite{McLerran:2007qj} that suggests a similar structure for QCD.
The investigations of these phases is however limited, mainly because they are technically much more involved.
Within the NJL model such phases have been analyzed at vanishing temperatures applying further truncations~\cite{Rapp:2000zd} as well as for the so-called chiral density wave~\cite{Sadzikowski:2000ap,Nakano:2004cd}. In the latter case the order parameter is assumed to be a plane wave and can be solved on mean-field level for vanishing current quark masses. In addition it has been suggested recently that the first order phase transition in the phase diagram of the NJL model, at least in the vicinity of the chiral critical point, is replaced by two second order phase transition lines that border an inhomogeneous phase~\cite{Nickel:2009ke}.
The intersection of the two lines defines a Lifshitz point, which coincides with the critical point in the NJL model.

\noindent
Inhomogeneous phases have also been investigated in the 1+1 dimensional (chiral) GN model\footnote{The chiral GN model is also called the 1+1 dimensional NJL model. The global  $Z_2$ symmetry of the GN model is extended to a $U(1)$ symmetry in this model.} for the large $N$ limit, which technically corresponds to a mean-field approximation.
Not only the understanding of the phase diagram has been improved significantly as it possesses regions with inhomogeneous  ground-states~\cite{Schnetz:2004vr,Schnetz:2005ih,Thies:2006ti}, also theoretical understanding of the integrability of these models has been obtained.
In the chiral limit all self-consistent solutions can in principle be classified~\cite{Correa:2009xa} and the phase diagrams have been analyzed in detail~\cite{Basar:2009fg}. For the case of the GN model it is furthermore possible to introduce finite quark masses and to study their effect on the structure of the phase diagram~\cite{Schnetz:2005ih}.

\noindent
The underlying idea of the present investigation is based on the observation that the theoretical problem of finding self-consistent inhomogeneous phases with a lower dimensional modulation can be reduced to a problem in a lower dimensional model. This is mainly due to the structure of the mean-field Hamiltonian, which is of Dirac-type.
As a consequence self-consistent phases with a one-dimensional modulation in the NJL model can be studied on similar grounds as in the 1+1 dimensional (chiral) GN model and all results obtained for the latter can be used here.
In particular all self-consistent inhomogeneous solutions with a one-dimensional modulation are in principle also known for the NJL model, at least in the chiral limit.
Furthermore we can investigate self-consistent inhomogeneous phases at finite quark masses, which has not been possible before.
The considered inhomogeneous ground states are then lattices of domain-wall solitons.
In the same way self-consistent phases with a two-dimensional modulation could be related to those in the 2+1 dimensional GN model.
However no analytical solutions for a real two-dimensional modulation are known for the latter.

\noindent
Being able to investigate the role of inhomogeneous phases in the phase diagram of the NJL model, we can confirm the picture obtained in Ref.~\cite{Nickel:2009ke} for the vicinity of the critical point and the absence of a first order phase transition line in the phase diagram\footnote{This picture could of course be modified by the inclusion of color-superconducting phases.}. Furthermore we can analyze the relation between inhomogeneous phases and e.g. the strength of the first order phase transition (present in the case of homogeneous phases).
For the main part of the paper we do not allow for pseudo-scalar condensates, especially since only without  those self-consistent solutions for finite current quark masses are known. We will however also discuss why this condensates are not expected for the ground-state.

\noindent
At the end of the paper we will finally also consider the QM model. The purpose of this is twofold: On the one hand we would like to extend the analysis to a larger class of models in general, on the other hand the regularization of the NJL model for inhomogeneous phases is non-trivial with regard to a combined vacuum and QCD phase diagram phenomenology.

\noindent
The paper is organized as follows:
In section~\ref{sec:NJL} we introduce the NJL model and the applied mean-field approximation allowing for inhomogeneous phases.
After this we show in section~\ref{sec:lowerdimmod} how self-consistent solutions for lower-dimensional modulations can be obtained from lower dimensional models.
We then summarize self-consistent solutions of the (chiral) GN model in section~\ref{sec:1d} and work out in section~\ref{sec:1dNJL} how those can be used for the NJL model.
Before coming to the QM model in section~\ref{sec:QM}, we remark in section~\ref{sec:pseudo} why pseudo-scalar condensates are not expected.
Finally we discuss various numerical results in section~\ref{sec:numerics} and close with a summary and outlook in section~\ref{sec:discussion}.

\section{NJL model and mean-field approximation}
\label{sec:NJL}
\noindent
In this work we concentrate on the two-flavor NJL model given by the Lagrangian
\bea
\mathcal{L}
&=&
\bar{\psi}
\left(
i\gamma^\mu \partial_\mu
-
\hat{m}
\right)
\psi
+
G_s
\left(
\left(\bar{\psi}\psi\right)^2
+
\left(\bar{\psi}i\gamma^5\tau^a\psi\right)^2
\right)
\,,
\eea
where $\psi$ is the $4N_f N_c$-dimensional quark spinor for $N_f=2$ flavors and $N_c=3$ colors, $\gamma^\mu$ are Dirac matrices, $G_s$ is the scalar coupling and $\hat{m}$ the mass matrix for degenerate quarks with current quark mass $m$. For $N_f=2$ the matrices $\tau^a$ are the conventional Pauli matrices.

\noindent
We address phases with non-vanishing expectation values $\langle\bar{\psi}\psi\rangle=S(\x)$ and  $\langle\bar{\psi}i\gamma^5\tau^a\psi\rangle=P_a(\x)$. Expanding bilinears around those expectation values and neglecting quadratic contributions, we then work within mean-field approximation. For technical reasons that will become more apparent in the following we furthermore restrict ourselves to the case where the direction of the vector $P_a(\x)$ is constant in flavor space so that we can choose a flavor basis where $P_1(\x)=P_2(\x)=0$ and $P_3(\x)=P(\x)$. The mean-field Lagrangian therefore takes the form
\bea
\label{eq:LMF}
\mathcal{L}_{MF}
&=&
\bar{\psi}(i\gamma^\mu \partial_\mu -m +2G_s S(\x) + 2iG_s\gamma^5 \tau^3 P(\x))\psi
-
G_s
\left(
S(\x)^2 +P(\x)^2
\right)
\,.
\eea
Since the mean-field Lagrangian is bilinear, the thermodynamic potential as an effective action in the expectation values can be formally evaluated.
In the case of a periodic condensate with Wigner-Seitz cell $V$ and using the imaginary-time formalism (see e.g. Refs.~\cite{reviews}), the thermodynamic potential as an effective action in $S(\x)$, $P(\x)$ is then given by
\bea
\label{eq:Omega1}
\Omega(T,\mu;S(\x),P(\x))
&=&
-\frac{T}{V}\ln 
\int \mathcal{D}\bar{\psi}\mathcal{D}\psi \exp\left(\int_{x\in [0,\frac{1}{T}]\times V} (\mathcal{L}_{MF}+\mu \bar{\psi}\gamma^0 \psi)\right)
\nonumber\\
&=&
-\frac{T N_c}{V}
\sum_{n}
\mathrm{Tr}_{D,f,V} \, \mathrm{Log}\left(\frac{1}{T}\left(i\omega_{n}+\tilde{H}_{MF}-\mu\right)\right)
+
\frac{G_s}{V}\int_V 
\left(
S(\x)^2
+
P(\x)^2
\right)
\,,
\nonumber\\
\eea
where the functional logarithm and trace act on Dirac, color and coordinate space. The \mbox{Hamiltonian} $\tilde{H}_{MF}$ is obtained from Eq.(\ref{eq:LMF}) and reads
\bea
\label{eq:hamiltonian0}
\tilde{H}_{MF}
&=&
-i\gamma^0\gamma^{i}\partial_{i} + \gamma^0\left(m - 2G_s S(\x) - 2iG_{s}\gamma^5\tau^3P(\x)\right)
\,,
\eea
which is a direct product of the two isospectral Hamiltonians
\bea
\label{eq:hamiltonian1}
H_{MF,\pm}
&=&
-i\gamma^0\gamma^{i}\partial_{i} + \gamma^0\left(m - 2G_s S(\x) \mp 2iG_{s}\gamma^5 P(\x)\right)
\,.
\eea
Assuming a sensible regularization for the functional trace we can then express the thermodynamic potential through the eigenvalues $\{E_n\}$  of $H_{MF,+}$ as
\bea
\label{eq:Omega2}
\Omega(T,\mu;M(\x))
&=&
-\frac{2T N_c}{V}
\sum_{E_{n}}
\ln\left(2\cosh\left(\frac{E_{n}-\mu}{2T}\right)\right)
+
\frac{1}{V}\int_V
\frac{\vert M(\x)-m\vert^2}{4G_s}
+
\text{const.}
\,,
\eea
where $M(\x) = m-2G_s(S(\x)+iP(\x))$.
We note that the mass function $M(\x)$ as the order parameter can be complex and parameterizes a $U(1)$ subgroup.

\noindent
In order to minimize the thermodynamic potential in the order parameter $M(\x)$ we can address the stationary constraint $\frac{\delta\Omega}{\delta M(\x)^*}=0$, which in turn gives the gap-equation
\bea
\label{eq:gap1}
M(\x)
&=&
m
+
\frac{4G_{s}N_{c}}{V}
\sum_{E_{n}}
\tanh\left(\frac{E_{n}-\mu}{2T}\right)
\psi^\dagger_{n}(\x)
\frac{\partial H_{MF,+}}{\partial M(\x)^*}
\psi_{n}(\x)
\nonumber\\
&=&
m
+
\frac{2G_{s}N_{c}}{V}
\sum_{E_{n}}
\tanh\left(\frac{E_{n}-\mu}{2T}\right)
\bar{\psi}_{n}(\x)
\left(1-\gamma^5\right)
\psi_{n}(\x)
\eea
as a self-consistency condition on $M(\x)$.
Here $\psi_{n}(\x)$ are the normalized eigenvectors of the Hamiltonian for the eigenvalues $E_{n}$, i.e. $\frac{1}{V}\int_V\psi^\dagger_{n}\psi_{n}=1$.

\section{lower-dimensional modulations}
\label{sec:lowerdimmod}
\noindent
Having the formal expressions for thermodynamic potential and gap-equation, we turn towards inhomogeneous phases. The focus is the determination of the eigensystem of a Hamiltonian and the check of self-consistency.
As will be discussed in the following, these problems can be dimensionally reduced for lower-dimensional modulations by help of Lorentz symmetry.

\noindent
Suppose we have a representation $S(\Lambda)$ for the Lorentz transformation $\Lambda$ that acts on
the momentum operator of free spinors $P^{\mu}=(H,P_i)^\mu$ as
\bea
S(\Lambda)P^\mu S^{-1}(\Lambda)
&=&
\Lambda^{\mu}_{\phantom{\mu}\nu}P^\nu
\,.
\eea
If the system is translationally invariant in one or more directions, the corresponding momenta $P_\perp$ commute with the Hamiltonian $H$ and we can label its eigenstates also by $\p_\perp$, the eigenvalue of $P_\perp$. Now take $\psi_{\lambda,\0}$ to be the eigenvector with $H\psi_{\lambda,\0}=\lambda\psi_{\lambda,\0}$ and  $P_\perp\psi_{\lambda,\0}=0$,
$\Lambda^{\mu}_{\phantom{\mu}\nu}$ to be the Lorentz transformation that boosts $(\lambda,\0)^\mu$ to $\left(\lambda\sqrt{1+\p_\perp^2/\lambda^2},\p_\perp\right)^\mu$ and define
\bea
\psi_{\lambda\sqrt{1+\p_\perp^2/\lambda^2},\p_\perp}
&=&
\left(\sqrt{1+\p_\perp^2/\lambda^2}\right)^{-\frac{1}{2}}S^{-1}(\Lambda)\psi_{\lambda,\0}
\,.
\eea
The prefactor implements the proper normalization for the conventional choice of $S(\Lambda)$~\cite{Itzykson:1980rh} and it is straightforward to check that
\bea
P^{\mu} 
\psi_{\lambda\sqrt{1+\p_\perp^2/\lambda^2},\p_\perp}
&=&
\left(\lambda\sqrt{1+\p_\perp^2/\lambda^2},\p_\perp\right)^\mu
\psi_{\lambda\sqrt{1+\p_\perp^2/\lambda^2},\p_\perp}
\,.
\eea
Therefore the whole eigenvalue spectrum can be constructed from the subspace spanned by $\{\psi_{\lambda,\0}\}$ and
for the thermodynamic potential we obtain
\bea
\label{eq:Omega3}
\Omega(T,\mu;M(\x))
&=&
-\frac{2T N_c}{V_\parallel}
\sum_{\lambda}
\int\frac{d\p_\perp}{(2\pi)^{d_\perp}}
\ln\left(2\cosh\left(\frac{\lambda\sqrt{1+\p_\perp^2/\lambda^2}-\mu}{2T}\right)\right)
\nonumber\\&&
+
\frac{1}{V}\int_V
\frac{\vert M(\x)-m\vert^2}{4G_s}
+
\text{const.}
\,,
\eea
where $V_\parallel$ is a Wigner-Seitz cell of the lower dimensional modulation.
Addressing the self-consistency condition in Eq.(\ref{eq:gap1}) we note that $S(\x)$ and $P(\x)$ are scalar and pseudo-scalar, respectively. We find
\bea
\bar{\psi}_{\lambda\sqrt{1+\p_\perp^2/\lambda^2},\p_\perp}
\psi_{\lambda\sqrt{1+\p_\perp^2/\lambda^2},\p_\perp}
&=&
\frac{\lambda}{\sqrt{\lambda^2+\p_\perp^2}}
\bar{\psi}_{\lambda,\0}
\psi_{\lambda,\0}
\,,
\nonumber\\
\bar{\psi}_{\lambda\sqrt{1+\p_\perp^2/\lambda^2},\p_\perp}
\gamma^5
\psi_{\lambda\sqrt{1+\p_\perp^2/\lambda^2},\p_\perp}
&=&
\frac{\lambda}{\sqrt{\lambda^2+\p_\perp^2}}
\bar{\psi}_{\lambda,\0}
\gamma^5
\psi_{\lambda,\0}
\eea
and the gap-equation takes the form
\bea
\label{eq:gap2}
M(\x)
&=&
m
+
\frac{2G_{s}N_{c}}{V_\parallel}
\sum_{\lambda}
\int\frac{d\p_\perp}{(2\pi)^{d_\perp}}
\tanh\left(\frac{\lambda\sqrt{1+\p_\perp^2/\lambda^2} -\mu}{2T}\right)
\frac{\lambda}{\sqrt{\lambda^2+\p_\perp^2}}
\bar{\psi}_{\lambda,\0}
\left(1-\gamma^5\right)
\psi_{\lambda,\0}
\,.
\nonumber\\
\eea

\noindent
Therefore we have formal expressions for the thermodynamic potential and the gap-equation in terms of the subspace spanned by the eigensystem corresponding to $\p_\perp=0$.
To be more specific we now 
choose the Weyl representation for the $\gamma$-matrices, i.e.
\bea
\gamma^0
\,
=
\,
\left(
\begin{array}{cc}
0 & \mathbf{1}\\
\mathbf{1} & 0
\end{array}
\right)
\,,
\quad
\gamma^i
\,
=
\,
\left(
\begin{array}{cc}
0 &  \sigma^i \\
-\sigma^i & 0
\end{array}
\right)
\,,
\quad
\gamma^5
\,
=
\,
\left(
\begin{array}{cc}
-\mathbf{1}&0\\
0&\mathbf{1}
\end{array}
\right)
\eea
with the conventional Pauli matrices $\sigma^i$ so that the Hamiltonian in Eq.(\ref{eq:hamiltonian1}) takes the form
\bea
\label{eq:hamiltoniandirac}
H_{MF,+}
&=&
\left(
\begin{array}{cc}
i\sigma^i \partial_i & M(\x)\\
M(\x)^* & -i\sigma^i \partial_i
\end{array}
\right)
\,.
\eea

\noindent
First restricting to a one-dimensional modulation in the $z$-direction, the Hamiltonian in the subspace corresponding to $\p_\perp = 0$ is
\bea
H_{MF;1D}
&=&
\left(
\begin{array}{cccc}
i\partial_z & & M(z)\\
&-i\partial_z && M(z)\\
M(z)^* & &-i\partial_z&\\
&M(z)^* & &i\partial_z
\end{array}
\right)
\,,
\eea
which can be block-diagonalized into
\bea
H'_{MF;1D}
&=&
\left(
\begin{array}{cc}
H_{1D}(M(z)) & \\
& H_{1D}(M(z)^*)
\end{array}
\right)
\,,
\eea
where
\bea
\label{eq:H1D}
H_{1D}(M(z))
&=&
\left(
\begin{array}{cc}
-i\partial_z & M(z)\\
M(z)^*& \phantom{-}i\partial_z
\end{array}
\right)
\,.
\eea
The problem is therefore reduced to finding the quasi-particle spectrum of $H_{1D}$ which is the Hamilton of the chiral GN model and which will be discussed in more detail in section~\ref{sec:1d}.

\noindent
Considering the scenario with a two dimensional modulation it is more convenient to assume the system to be translationally invariant in the $z$-direction. The Hamiltonian in Eq.(\ref{eq:hamiltoniandirac}) for $\p_\perp = 0$ is then given by
\bea
H_{MF;2D}
&=&
\left(
\begin{array}{cccc}
 & i\partial_x-\partial_y& M(x,y)\\
i\partial_x+\partial_y& && M(x,y)\\
M(x,y)^* & &&-i\partial_x+\partial_y\\
&M(x,y)^* &-i\partial_x-\partial_y &
\end{array}
\right)
\,.
\eea
This can be block-diagonalized only for real $M(x,y)$, i.e. for $P(\x)=0$. In this case we can cast the problem into the form
\bea
\tilde{H}_{MF;2D}
&=&
\left(
\begin{array}{cccc}
-M(x,y) & i\partial_x+\partial_y& \\
i\partial_x-\partial_y&-M(x,y) && \\
 & &M(x,y)&i\partial_x+\partial_y\\
& &i\partial_x-\partial_y &M(x,y)
\end{array}
\right)
\,,
\eea
which corresponds to the GN model in two-dimensions. The fact that we need a real mass function $M(x,y)$ is not surprising since there is no chiral GN model with two-component spinors in two dimensions.

\noindent
In general we are therefore able to constrain ourselves to a lower dimensional model if we want to determine the quasi-particle spectrum for an inhomogeneous phase with a lower dimensional modulation.
However a self-consistent modulation in the lower dimensional model does not necessarily have to be self-consistent in the higher dimensional model.
For the cases considered in this paper and essentially all analytical solutions known this will nevertheless be the case since for each eigenvector separately we will have
\bea
\label{eq:abM}
\bar{\psi}_{\lambda,\0}
\left(1-\gamma^5\right)
\psi_{\lambda,\0}
&\propto&
a + b M(z)
\eea
and there are enough parameters in the analytical expression for the modulation to tune the coefficients $a$, $b$ such that the solution is self-consistent.

\section{inhomogeneous phases in one-dimensional models}
\label{sec:1d}
\noindent
Having realized that self-consistent solutions with lower dimensional modulations may be obtained from lower dimensional models, we now present some self-consistent phases with a one-dimensional modulation. This reduces to the study of the (chiral) GN model whose Hamiltonian is given by Eq.(\ref{eq:H1D}) and in the following we summarize some results deduced from Refs.~\cite{Schnetz:2004vr,Schnetz:2005ih,Basar:2008im,Correa:2009xa}.

\noindent
The thermodynamic potential can be evaluated by the knowledge of the density of states $\rho(\lambda)$ which enables us to perform the sum over the eigenvalue spectrum: For an arbitrary function $f(\lambda)$ it is given by
\bea
\frac{1}{V_\parallel}
\sum_\lambda f(\lambda)
&=&
\int\! d\lambda\, \rho(\lambda) f(\lambda)
\,.
\eea
Recent investigations have shown that at least for the chiral limit all self-consistent solutions can be classified and that all information about the inhomogeneous phase is encoded in the band-edge energies $\{E_i\}$ of the valence bands of the quasi-particles~\cite{Correa:2009xa}.
For the case of one gapped valence band the general solution for a complex order parameter $M(z)$ has been determined~\cite{Basar:2008im} and the spectral density associated with the Hamiltonian $H_{1D}$ is given by
\bea
\label{eq:rho}
\rho(\lambda)
&=&
\left\{
\begin{array}{ll}
\frac{1}{2\pi}
\,
\frac{a(\lambda)+\langle\vert M(z)\vert^2\rangle}{\sqrt{\Pi_{j=1}^4 (\lambda-E_j)}}
&
\quad,\,\Pi_{j=1}^4 (\lambda-E_j) >0
\\
0 &\quad, \, \text{else}
\end{array}
\right.
\,,
\eea
where
\bea
a(\lambda)
&=&
2\Big(
\lambda
-
\frac{1}{4}
\sum_{j=1}^4 E_j
\Big)^2
-
\frac{1}{8}
\sum_{i<j} (E_i - E_j)^2
\,,\nonumber\\
\langle\vert M(z)\vert^2\rangle
&=&
\frac{1}{4}
(E_1+E_2-E_3-E_4)^2
-
\frac{
(E_1-E_3)(E_2-E_4) \E(\frac{(E_1-E_2)(E_3-E_4)}{(E_1-E_3)(E_2-E_4)})
}{
\K(\frac{(E_1-E_2)(E_3-E_4)}{(E_1-E_3) (E_2-E_4)})
}
\eea
and $E_1\leq E_2\leq E_3\leq E_4$ are the edges of the valence bands\footnote{The expression for $\langle\vert M(z)\vert^2\rangle$ has not been derived in Ref.~\cite{Basar:2008im}, but is obtained by solving the Ginburg-Landau equations and presenting the solution in terms of Jacobi elliptic functions. Since it is not relevant for the main focus of this work, we do not give the tedious derivation.}.
For more details  on the function $M(z)$ we refer to Refs.~\cite{Basar:2008im,Basar:2009fg} since we will limit ourselves to real order parameters $M(z)$ in the following.

\noindent
For real order parameters $M(z)$ we have $\rho(\lambda)=\rho(-\lambda)$ or equivalently $E_1=-E_4$, $E_2=-E_3$. So there are only two parameters to specify the solution, which we choose to be given through $E_3=\sqrt{1-\nu}\Delta$ and $E_4=\Delta$~\cite{Schnetz:2004vr}.
This case is of particular interest since it is possible to generalize the solution to finite current quark masses~\cite{Schnetz:2005ih}. The order parameter then has the specific form
\bea
\label{eq:MzReal}
M(z)
&=&
\Delta
\left(
\nu\,
\mathrm{sn}(b\vert \nu)
\mathrm{sn}(\Delta z\vert \nu)
\mathrm{sn}(\Delta z+b\vert \nu)
+
\frac{
\mathrm{cn}(b\vert \nu)\mathrm{dn}(b\vert \nu)
}{
\mathrm{sn}(b\vert \nu)
}
\right)
\,,
\eea
where $\Delta$ is a scale parameter and $\mathrm{sn}$, $\mathrm{cn}$, $\mathrm{dn}$ are elliptic Jacobi functions with elliptic modulus $\sqrt{\nu}$, which
physically describes lattices of equidistant solitions.
Due to periodicity the parameters $b$ and $\nu$ can be limited to $b\in [0,\K(\nu)]$ and $\nu\in [0,1]$ with the quarter period $\K(\nu)$.  The chiral limit, i.e. vanishing quark masses, corresponds to $b=\K(\nu)$.

\noindent
In terms of the eigenvalue spectrum, the case of finite current quark masses included through $b\neq\K(\nu)$ only leads to a shift in the distribution of the eigenvalues $\lambda$
\bea
\lambda
&\rightarrow&
\mathrm{sign}(\lambda)\sqrt{\lambda^2 + \delta \Delta^2}
\,,
\nonumber\\
\delta
&=&
\frac{
1
}{
\mathrm{sn}^2(b\vert \nu)
}
-1
\,\,
\geq
\,\, 0
\eea
and as a consequence the sum over the eigenvalue spectrum for arbitrary functions $f(\lambda)$ can be performed by using the density of states $\rho(\lambda)$ of the chiral limit
\bea
\frac{1}{V_\parallel}\sum_\lambda f(\lambda)
&=&
\int_{-\infty}^{\infty}\! d\lambda\, \rho(\lambda) f(\lambda)
\nonumber\\
&\stackrel{\mathrm{\delta\neq 0}}\longrightarrow&
\int_{-\infty}^{\infty}\! d\lambda\,\rho(\lambda)
f(\lambda\sqrt{1+\delta \Delta^2/\lambda^2})
\,.
\eea

\noindent
Addressing self-consistency and following Ref.~\cite{Schnetz:2005ih}, we have for the normalized eigenvectors $\psi_{1D}(z)$ of $H_{1D}$ 
\bea
\label{eq:psibarpsi}
\bar{\psi}_{1D}(z)\psi_{1D}(z)
&=&
\frac{\lambda}{\lambda^2-\delta\Delta^2-\Delta^2\E(\nu)/\K(\nu)}\,M(z)
-
\frac{\Delta^3\sqrt{\delta  (\delta +1) (\delta -\nu +1)}}{\lambda(\lambda^2-\delta\Delta^2-\Delta^2\E(\nu)/\K(\nu))}
\,.
\eea
As discussed in the context of Eq.(\ref{eq:abM}), we observe that the first term in Eq.(\ref{eq:psibarpsi}) is proportional to the modulation $M(z)$ whereas the second is constant for each eigenvector.
This enables us to obtain self-consistency in the gap-equation for a proper choice of parameters.
In particular we find that the second term vanishes for $\delta=0$ which then corresponds to the chiral limit.
Furthermore we have
\bea
\bar{\psi}_{\lambda,\0}
\left(1-\gamma^5\right)
\psi_{\lambda,\0}
&=&
\bar{\psi}_{1D}\psi_{1D}
\eea
and conclude with the same reasoning that we can achieve self-consistency for $M(z)$ being a one-dimensional modulation in the NJL model.

\section{One-dimensional modulations in the NJL model}
\label{sec:1dNJL}
\noindent
We will now concentrate on real order parameters $M(z)$ and parametrize the band edges by $E_3=\sqrt{1-\nu}\Delta$, $E_4=\Delta$, which corresponds to the order parameter in Eq.(\ref{eq:MzReal}) for $b=\K(\nu)$. After a non-trivial computation using Eq.(\ref{eq:rho}) we obtain the density of states
\bea
\label{eq:rhoreal}
\rho(\lambda)
&=&
\left\{
\begin{array}{ll}
\frac{1}{\pi}
\frac{\lambda ^2- \Delta ^2 \E(\nu)/\K(\nu)
}{
\sqrt{\left(\lambda ^2-\Delta ^2\right) \left(\lambda ^2-(1-\nu) \Delta^2\right)}
}
&
\quad, \, \left(\lambda ^2-\Delta ^2\right) \left(\lambda ^2-(1-\nu) \Delta^2\right)>0
\\
0
&
\quad, \, \text{else}
\end{array}
\right.
\,,
\eea
which of course agrees with the finding in Ref.~\cite{Schnetz:2004vr}.
The evaluation of the thermodynamic potential in Eq.(\ref{eq:Omega3}) then reduces to evaluating expressions of the form
\bea
\label{eq:NJLsum}
\frac{1}{V_\parallel}
\sum_{\lambda}
\int\frac{d^2\p_\perp}{(2\pi)^{2}}
f\left(\lambda\sqrt{1+\p_\perp^2/\lambda^2}\right)
&\rightarrow&
2\int_{-\infty}^{\infty}\! d\lambda\,
\int\frac{d^2\p_\perp}{(2\pi)^{2}}\,
\rho(\lambda)
f\left(\mathrm{sign}(\lambda)\sqrt{\lambda^2+\p_\perp^2+\delta \Delta^2}\right)
\nonumber\\
&=&
\int_{-\infty}^{\infty}\! d\lambda\,
\int\frac{d^2\p_\perp}{(2\pi)^{2}}\,
\rho(\lambda)
\tilde{f}\left(\sqrt{\lambda^2+\p_\perp^2+\delta \Delta^2}\right)
\nonumber\\
&=&
\frac{1}{2\pi}
\int_{0}^{\infty}\! dE\,E^2
\int_{-1}^1 du\,
\rho(E u)
\tilde{f}\left(\sqrt{E^2+\delta \Delta^2}\right)
\nonumber\\
&=&
\int_{0}^{\infty}\! dE\,
\tilde{\rho}(E)
\tilde{f}\left(\sqrt{E^2+\delta \Delta^2}\right)
\,.
\eea
Here we introduced $\tilde{f}(x)=f(x)+f(-x)$, used $\rho(\lambda)=\rho(-\lambda)$ and implicitly defined the effective density of states
\bea
\tilde{\rho}(E)
&=&
\frac{1}{2\pi}
\int_{-1}^1 du\,E^2
\rho(E u)
\,.
\eea
The overall factor of two in Eq.(\ref{eq:NJLsum}) stems from the degeneracy of the two Hamiltonians $H_{1D}(M)$ and $H_{1D}(M^*)$ in $H_{MF;1D}$.
A straightforward but somewhat tedious calculation using the density of states in Eq.(\ref{eq:rhoreal}) then gives the explicit expression
\bea
\tilde{\rho}(E)
&=&
\phantom{+}
\theta(\sqrt{\tilde{\nu}}\Delta-E)
\frac{
 \E(\tilde{\theta} |\tilde{\nu})+(\E(\nu)/\K(\nu)-1) \F(\tilde{\theta} |\tilde{\nu})
}{
\pi^2
}
E\Delta
\nonumber\\&&
+
\theta(E-\sqrt{\tilde{\nu}}\Delta)
\theta(\Delta-E)
\frac{
 \E(\tilde{\nu})+(\E(\nu)/\K(\nu)-1) \K(\tilde{\nu})
}{
\pi^2
}
E\Delta
\nonumber\\&&
+
\theta(E-\Delta)
\frac{
\E(\theta |\tilde{\nu})
+\E(\nu)/\K(\nu) \F(\theta |\tilde{\nu})
-\F(\theta |\tilde{\nu})
+\sqrt{(E^2 - \Delta^2) (E^2 - \tilde{\nu}\Delta^2)}/(E\Delta)
}{\pi ^2}
E\Delta
\,,
\nonumber\\
\eea
where we introduced the elliptic integrals of 1st and 2nd kind as well as
$\tilde{\nu}=1-\nu$, $\tilde{\theta}=\arcsin(E/(\sqrt{\tilde{\nu}}\Delta))$ and $\theta=\arcsin(\Delta/E)$.
As an interesting cross-check we can consider the limit
\bea
\left.\tilde{\rho}(E)\right|_{\nu=1}
&=&
\phantom{+}
\frac{
1
}{\pi ^2}
\theta(E-\Delta)
\sqrt{(E^2 - \Delta^2)}
E
\,,
\eea
which is the effective density of states in an homogeneous phase with a quasiparticle gap $\Delta$ and
\bea
\left.\tilde{\rho}(E)\right|_{\nu=0}
&=&
\frac{1}{\pi^2}
E^2
\,,
\eea
which corresponds to the ultra-relativistic gas. We also find for the asymptotic behavior of the effective density of states
\bea
\tilde{\rho}(E)
&=&
\frac{E^2}{\pi ^2}
-
\frac{\langle M(z)^2\rangle}{2 \pi^2}
-
\frac{
\langle M(z)^4\rangle
+
\langle M'(z)^2\rangle
}{8 \pi ^2 E^2}
+
O\Big(\left(\frac{1}{E}\right)^4\Big)
\,.
\eea

\noindent
For the considered one-dimensional modulation the thermodynamic potential in Eq.(\ref{eq:Omega3}) can therefore be cast into
\bea
\label{eq:OmegaNJLNum}
\Omega_{MF,NJL}(T,\mu;\Delta,\nu,\delta)
&=&
-
2N_c
\int_0^\infty\!dE\,
\tilde{\rho}(E)
\tilde{f}_{\text{bare}}\left(\sqrt{E^2+\delta\Delta^2}\right)
+
\frac{1}{4G_sL}\int_0^L
\!dz\,
\vert M(z)-m\vert^2
\nonumber\\&&
+
\text{const.}
\,,
\phantom{aa}
\eea
where $L$ is the period of the modulation and
\bea
\label{eq:fx}
\tilde{f}_{\text{bare}}(x)
&=&
T\ln\left(2\cosh\left(\frac{x-\mu}{2T}\right)\right)
+
T\ln\left(2\cosh\left(\frac{x+\mu}{2T}\right)\right)
\nonumber\\
&=&
\tilde{f}_{\text{UV}}(x)
+
\tilde{f}_{\text{medium}}(x)
\,,
\nonumber\\
\tilde{f}_{\text{UV}}(x)
&=&
x
\,,
\nonumber\\
\tilde{f}_{\text{medium}}(x)
&=&
T
\ln\left(
1+\exp\left(-\frac{x-\mu}{T}\right)
\right)
+
T
\ln\left(
1+\exp\left(-\frac{x+\mu}{T}\right)
\right)
\,.
\eea
The last missing step is then a regularization of the diverging integration. For homogeneous phases this is mostly done by a momentum regularization (see e.g. Refs.~\cite{Scavenius:2000qd, reviews}). This is however not possible for inhomogeneous phases since the quasi-particle energies can no longer be labelled by a conserved three-momentum. Instead we have to apply a regularization of the functional logarithm, e.g. by a proper-time regularization, which is essentially a regularization acting on the energy spectrum instead of the quasi-particle momenta.
For this purpose we have already identified the divergent vacuum contribution associated to $\tilde{f}_{\text{UV}}(x)$, which we regularize by a specific blocking function in the proper-time integral leading to a Pauli-Villars regularization of the form~\cite{Klevansky:1992qe}
\bea
\tilde{f}_{\text{UV}}(x)
&\rightarrow&
\tilde{f}_{\text{PV}}(x)
\quad=\quad
\sum_{j=0}^{3}c_j\sqrt{x^2+j\Lambda^2}
\,,
\eea
with $c_0=1$, $c_1=-3$, $c_2=3$, $c_3=-1$ and a cutoff scale $\Lambda$.

\noindent
With the expression of the thermodynamic potential as a function of $\Delta$, $\nu$ and $\delta$ we are now able to perform numerical investigations by performing the energy integration numerically and minimizing in the mentioned parameters. The resulting phase diagrams will of course depend on the choice of the model parameters $G_s$, $\Lambda$ and $m$.
Those can be related to the chiral condensate $\langle\bar{\psi}\psi\rangle$ and the pion decay constant though the expressions~\cite{Klevansky:1992qe}
\bea
\label{eq:cbarcfpi}
\langle\bar{\psi}\psi\rangle
&=&
-\frac{3M_q}{4\pi^2}\sum_{j=0}^{3}c_j(M_q^2+j\Lambda^2)\log\left(\frac{M_q^2+j\Lambda^2}{M_q^2}\right)
\,,
\nonumber\\
f_\pi^2
&=&
-\frac{N_c M_q^2}{4\pi^2}\sum_{j=0}^{3}c_j\log\left(\frac{M_q^2+j\Lambda^2}{M_q^2}\right)
\,,
\eea
where $M_q$ is the constituent mass for the chirally broken phase in the vacuum.

\section{A note on pseudo-scalar condensates}
\label{sec:pseudo}
\noindent
Having worked out an expression for the thermodynamic potential for real order parameters $M(z)$, we may also discuss the more general case of a complex order parameter or in other words the relevance of pseudo-scalar condensates $P_a(x)$. 

\noindent
On the one hand side this is less attractive from a technical point of view, since we only know self-consistent solutions away from the chiral limit for real order parameters $M(z)$.
More importantly we can also argue that these phases are energetically less preferred, at least in the vicinity of a second order phase transition to the chirally restored phase and in the chiral limit.

\noindent
For this purpose we consider the generalized Ginzburg-Landau (GL) functional which is a systematic expansion of the thermodynamic potential in the magnitude of the order parameter $M(z)$ as well as in gradients acting on it. Both are treated to be of the same order. For the chiral GN model this has been worked out in Ref.~\cite{Boehmer:2007ea}, where
\bea
\label{eq:GLGN}
\Omega_{GL,GN}(M)-\Omega_{GL,GN}(0)
&=&
\phantom{+}\frac{\alpha_2}{2} \vert M\vert^2
+
\frac{\alpha_3}{3} \mathrm{Im}(M M'^{*})
+
\frac{\alpha_4}{4} (\vert M\vert^4 + \vert M'\vert^2)
\nonumber\\
&&
+
\frac{\alpha_5}{5} \mathrm{Im}((M'' - 3 \vert M\vert^2 M)M'^{*})
\nonumber\\
&&
+
\frac{\alpha_6}{6}(\vert M\vert^6 + 3 \vert M \vert^2 \vert M' \vert^2+ 2 \vert M\vert^{2} \vert M^2\vert'+ \frac{1}{2}\vert M'' \vert^2)
+\dots
\eea
has been derived. The dependence on temperature and chemical potential is hidden in the GL coefficients $\alpha_i$ and the details are irrelevant for the present discussion. 
For the NJL model the GL expansion for real order parameters has been worked out in Ref.~\cite{Nickel:2009ke} and a remarkable similarity to the GN model has been found when restricting to one-dimensional modulations.
From the viewpoint of Lorentz symmetry as presented in section~\ref{sec:lowerdimmod} this may not be surprising since the GL coefficients are defined as integrals over functions in momentum space, so that at least for $\p_\perp=0$ the integrands of chiral GN and NJL model have to be equal up to an overall prefactor. Since the rotationally invariant extension to $\p_\perp\neq0$ is unique, the structure of the GL expansions has to be the same.

\noindent
In contrast to the chiral GN model the NJL Hamiltonian is however the direct product of $H_{1D}(M)$ and $H_{1D}(M^*)$. Therefore the structure of the GL functional for the NJL model limited to one-dimensional modulations corresponds to the sum of that in Eq.(\ref{eq:GLGN}) and its conjugate. We therefore obtain
\bea
\Omega_{GL,NJL}(M)-\Omega_{GL,NJL}(0)
&=&
\phantom{+}
\frac{\beta_2}{2} \vert M\vert^2
+
\frac{\beta_4}{4} (\vert M\vert^4 + \vert M'\vert^2)
\nonumber\\&&
+
\frac{\beta_6}{6}(\vert M\vert^6 + 3 \vert M \vert^2 \vert M' \vert^2+ 2 \vert M\vert^{2} \vert M^2\vert'+ \frac{1}{2}\vert M'' \vert^2)
+\dots
\,,
\eea
where $\beta_i$ are the GL coefficients of the NJL model as stated in Ref.\cite{Nickel:2009ke}.

\noindent
With the generic form of the GL functional at hand, we can now compare various ans\"atze for $M(z)$. We will limit ourselves to second order phase transitions from an inhomogeneous to the chirally restored phase and therefore to a regime where $\beta_2>0$, $\beta_4<0$.
Furthermore all Fourier modes of $M(z)$ decouple to order $M^2$ due to momentum conservation in the GL coefficients, which are calculated in an homogeneous background. For this reason we consider only the two extremes
\bea
M_{FF}(z)
&=&
\frac{\Delta_{FF}}{\sqrt{2}}\exp(iqz)
\,,
\nonumber\\
M_{sin}(z)
&=&
\Delta_{sin}\sin(qz).
\eea
For an individually optimized wave-vector $q$ the thermodynamic potentials for these ans\"atze is then given by
\bea
\Omega_{GL,NJL}(M_{FF})-\Omega_{GL,NJL}(0)
&=&
\left(\frac{\beta _2}{4}-\frac{3 \beta _4^2}{32 \beta _6}\right) \Delta_{FF}^2-\frac{1}{8} \beta _4 \Delta _{FF}^4 + O(\Delta_{FF}^5)
\,,\nonumber\\
\Omega_{GL,NJL}(M_{sin})-\Omega_{GL,NJL}(0)
&=&
\left(\frac{\beta _2}{4}-\frac{3 \beta _4^2}{32 \beta _6}\right) \Delta_{sin}^2-\frac{1}{16} \beta _4 \Delta _{sin}^4 + O(\Delta_{sin}^5)
\,.
\eea
From this result we can deduce that the phase transition for both cases happens at the same values of the GL coefficients, namely at 
$\beta_2-\frac{3 \beta _4^2}{8 \beta _6}=0$. The inhomogeneous phase then occurs for $\beta_2-\frac{3 \beta _4^2}{8 \beta _6}<0$, where the coefficient of the $\Delta_{FF}^2$- and $\Delta_{sin}^2$-term is negative. Since the coefficient of the $\Delta_{sin}^4$-term is however smaller than that of the $\Delta_{FF}^4$-term (recall that $\beta_4<0$), we find that the modulation of the form $M_{sin}(z)$ is energetically preferred compared to $M_{FF}(z)$ for optimized parameters.
In this sense the ground state with a real order parameters wins over the plane wave. Two brief comments are in order here: First, the self-consistent modulations discussed in the previous section indeed become sinusoidal when approaching the phase transition to the chirally restored phase and second, the sinusoidal modulation can of course be understood as two plane waves in opposite directions. The latter have also been found energetically preferred compared to the plane wave in inhomogeneous (color-)superconductors~\cite{LO64,Bowers:2002xr}.

\section{QM model and mean-field approximation}
\label{sec:QM}
\noindent
The NJL model regularized by a proper-time regularization and adjusted to chiral condensate and pion decay constant is known to give constituent quark masses of order $200\mathrm{MeV}$ in the vacuum. Hence it gives an undesired phenomenology with regard to the QCD phase diagram, mainly because quasi-particles will start forming a Fermi surface at $\mu\simeq M_q$. This is phenomenologically unacceptable for $\mu< (M_N-B)/3\simeq308\mathrm{MeV}$, where $M_N$ is the nucleon mass and $B$ the binding energy of nucleons in nuclear matter.
In contrast the NJL model regularized by a sharp three-momentum cutoff gives constituent quark masses in the vacuum significantly larger than $300\mathrm{MeV}$.
Since we haven't found a regularization that avoids these problems, we instead also introduce a model that is very similar to the NJL model and where the issue of the regularization scheme can be surpassed: The linear sigma model, which in this context is usually named QM model~\cite{Scavenius:2000qd,Schaefer:2006ds}.

\noindent
The Lagrangian of the QM model with $N_f=2$ and $N_c=3$ is given by
\bea
\mathcal{L}_{QM}
&=&
\bar{\psi}
\left(
i\gamma^\mu \partial_\mu
-
g(\sigma+i\gamma_5 \tau^a\pi^{a})
\right)
\psi
-
U(\sigma,\pi^{a})
\,,
\eea
where
\bea
U(\sigma,\pi^{a})
&=&
-
\frac{1}{2}
\left(
\partial_\mu\sigma \partial^\mu\sigma 
+
\partial_\mu\pi^{a} \partial^\mu\pi^{a}
\right)
+
\frac{\lambda}{4}
\left(
\sigma^2
+
\pi^{a}\pi^{a}
-
v^2
\right)^{2}
-
c\sigma
\,,
\eea
$\psi$ is again the $4N_f N_c$-dimensional quark spinor, $\sigma$ the scalar field of the $\sigma$-meson and $\pi^{a}$ the pseudo-scalar fields of the pion triplet.
In mean-field approximation we treat the fields $\sigma$ and $\pi^{a}$ as classical and replace them by there expectation values~\cite{Scavenius:2000qd,Schaefer:2006ds}.
Furthermore we can use low-energy relations to connect the parameters $c$, $g$, $\lambda$ and $v^2$ with hadronic observables.
We will express those by the pion-decay constant $f_\pi$, the constituent quark mass in the vacuum $M_q$, the pion mass $m_\pi$ and $\sigma$-meson mass $m_\sigma$ via
$\langle\sigma\rangle=f_\pi$,
$\langle\pi^a\rangle=0$,
$c=m_\pi^2 f_\pi$,
$g=M_q/ f_\pi$,
$\lambda=(m_\sigma^2 - m_\pi^2)^2/ (2f_\pi^2)$ and
$v^2=f_\pi^2 - m_\pi^2/ \lambda$.

\noindent
For the thermodynamic potential in mean-field approximation we only include the contributions of the fermionic fluctuations and approximate
\bea
\label{eq:Omega1QM}
\Omega_{QM}(T,\mu;\sigma(\x),\pi^a(\x))
&=&
-\frac{T}{V}\ln 
\int \mathcal{D}\bar{\psi}\mathcal{D}\psi\mathcal{D}\sigma\mathcal{D}\pi^a
\exp\left(\int_{x\in [0,\frac{1}{T}]\times V} (\mathcal{L}_{QM}+\mu \bar{\psi}\gamma^0 \psi)\right)
\nonumber\\
&\stackrel{\text{mean-field}}\rightarrow&
-
\frac{T N_c}{V}
\sum_{n}
\mathrm{Tr}_{D,f,V} \, \mathrm{Log}\left(\frac{1}{T}\left(i\omega_{n}+\tilde{H}_{MF,QM}-\mu\right)\right)
\nonumber\\&&
+
\frac{1}{V}\int_V
U(\sigma(\x),\pi^a(\x))
\,,
\eea
where $\sigma(\x)$ and $\pi^a(\x)$ are non-vanishing expectation values of the respective fields.
Limiting to cases with $\pi^1(\x)=\pi^2(\x)=0$ the Hamiltonian reads
\bea
\label{eq:hamiltonian0QM}
\tilde{H}_{MF,QM}
&=&
-i\gamma^0\gamma^{i}\partial_{i} + \gamma^0\left(g\sigma(\x)+ig\gamma^5\tau^3\pi^3(\x)\right)
\,,
\eea
and we have to evaluate the same functional trace-logarithm as in the case of the NJL model, but now with the identification $M(\x)=g(\sigma(\x)+i\pi^3(\x))$.

\noindent
The QM model is renormalizable which means that the divergences in the functional trace-logarithm can be absorbed by the model parameters. For this purpose it is useful to observe that we can always separate zero-point fluctuations from thermal fluctuations as indicated by 
$\tilde{f}_{\text{UV}}(x)$ and
$\tilde{f}_{\text{medium}}(x)$
in Eq.(\ref{eq:fx}), respectively.
Instead of a proper renormalization we will however follow Refs.~\cite{Scavenius:2000qd,Schaefer:2006ds}, where it has been assumed that the  contribution from zero-point approximations can well be approximated by $\frac{1}{V}\int_V U(\sigma(\x),\pi^a(\x))$ with the parameters directly adopted to pheno\-menology.
As a result we can evaluate the thermodynamic potential for one-dimensional modulations $\sigma(\x)=M(z)/g$ and $\pi^a(\x)=0$ with $M(z)$ given in Eq.(\ref{eq:MzReal}) on the same level as for the NJL model and we get
\bea
\label{eq:OmegaNumQM}
\Omega_{MF,QM}(T,\mu;\Delta,\nu,\delta)
&\equiv&
\Omega_{MF,QM}(T,\mu;M(z)/g,0)
\nonumber\\&=&
-
2N_c
\int_0^\infty\!dE\,
\tilde{\rho}(E)
\tilde{f}_{\text{medium}}\left(\sqrt{E^2+\delta\Delta^2}\right)
+
\frac{1}{L}\int_0^L
\!dz\,
U(M(z)/g,0)
\nonumber\\&&
+
\text{const.}
\,.
\eea
For completeness we have to denote that this will only be a self-consistent solution if $M(z)$ would solve the equations of motions associated to minimizing the functional $U(\sigma(\x),\pi^a(\x))$, which is in general not the case.
This is not considered a major issue  within this work since we already limited the space of possible solutions to that of one-dimensional modulations so that we cannot guarantee to find the true ground state anyway. Instead we will find domains where at least one inhomogeneous phase is energetically preferred over all homogeneous phases and therefore pointing out domains where the ground-state is for sure inhomogeneous (in the presented approximation).

\noindent
At the end of this section we like to point out that we can easily analyze the vicinity of the chiral critical point by a generalized Ginzburg-Landau approximation. For this purpose we first observe that the  functional trace-logarithm in Eq.(\ref{eq:Omega1QM}) is formally identical to the the one in Eq.(\ref{eq:Omega1}). The GL expansion of the latter has been worked out in Ref.~\cite{Nickel:2009ke} and the structure of the GL functional stays the same when limiting to thermal fluctuations\footnote{The GL coefficients are defined through integrals in momentum space for which we can separate vacuum and thermal contributions.}.
The GL expansion of $\Omega_{MF,QM}(T,\mu;\sigma(\x)=M(\x)/g,0)$ can therefore easily deduced from Ref.~\cite{Nickel:2009ke}. Without giving unnecessary details for the presented discussion, the GL expansion with coefficients $\beta'_i$ takes the form
\bea
\Omega_{MF,QM}(T,\mu;\sigma(\x)=M(\x)/g,0)
&=&
\phantom{+}\frac{1}{V}
\int_V
\Big(
\phantom{+}
\frac{\beta'_{2}}{2}M(\x)^2
+
\frac{\beta'_4}{4}(M(\x)^4+(\nabla M(\x))^2)
\nonumber\\&&
\hspace{1.5cm}
+
\frac{\beta'_6}{6}(M(\x)^6+5M(\x)^2(\nabla M(\x))^2+(\Delta M(\x))^2)
\nonumber\\&&
\hspace{1.5cm}
+
\frac{v^4\lambda}{4}
-
cgM(\x)
-
\frac{g^2 v^2\lambda}{2}M(\x)^2
\nonumber\\&&
\hspace{1.5cm}
+
\frac{1}{4}(g^2\lambda M(\x)^4+2g^2(\nabla M(\x))^2)
\Big)
+\dots
\,.
\eea
Restricting to the chiral limit for simplicity, i.e. $c=0$, we can identify the chiral critical point as the position where the coefficients of the $M(\x)^2$-term and the $M(\x)^4$-term vanish, i.e.
\bea
\beta'_{2}-g^2v^2\lambda
&=&
0
\,,
\nonumber\\
\beta'_{4}+g^2\lambda
&=&
0
\,.
\eea
Considering the coefficient of the $(\nabla M(\x))^2$-term at this point we find
\bea
\frac{\beta'_{4}+g^2}{4}
&=&
\frac{g^2(1-\lambda)}{4}
\,,
\eea
being negative for $\lambda>1$.
Since homogeneous phases are instable for a vanishing $M(\x)^2$-term and a negative $(\nabla M(\x))^2$-term,
the chiral critical point would be hidden in this case since an inhomogeneous phase were energetically preferred at its location.
This scenario is actually realized for phenomenological parameters since we have $\lambda= m_\sigma^2/(2f_\pi^2)\sim 20$.
Due to the involved approximations we do however not want to overemphasize this result.

\section{numerical results}
\label{sec:numerics}

\begin{figure}
\includegraphics[width=7.5cm]{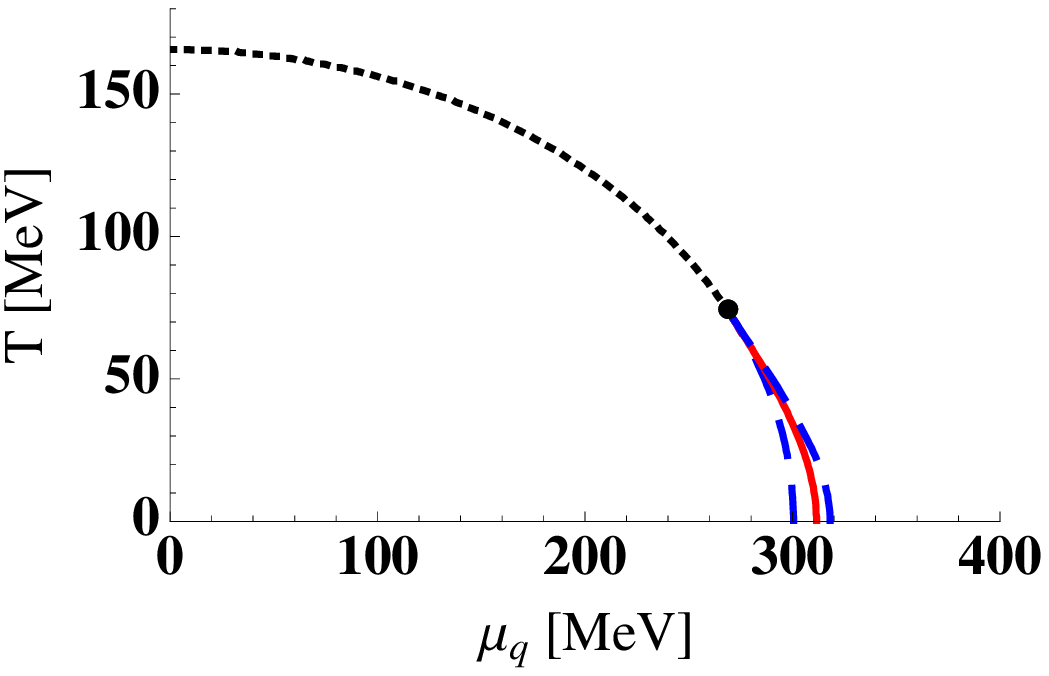}
\hspace{1cm}
\includegraphics[width=7.5cm]{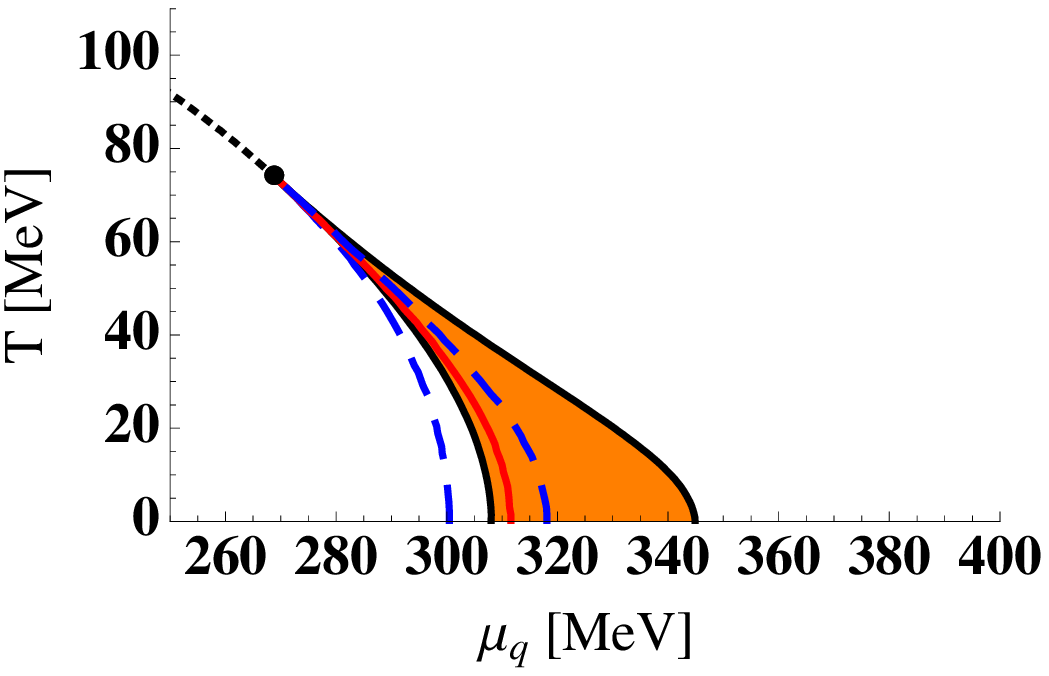}
\caption{
Left:
Structure of the NJL phase diagram in the chiral limit as a function of temperature $T$ and quark chemical potential $\mu_q$ for $M_q=300\mathrm{MeV}$.
The black (short-dashed) line indicates the second order phase transition from chirally broken to restored phase, the red (solid) line the first order phase transition and the bullet the critical point.
The spinodal region is enclosed by the blue (long-dashed) lines.
Right:
Same plot as on the left including the orange (shaded) domain where the energetically preferred ground state is inhomogeneous.
}
\label{fig1}
\end{figure}

\noindent
With the expressions for the thermodynamic potentials of NJL and  QM model in  Eq.(\ref{eq:OmegaNJLNum}) (combined with the described regularization) and Eq.(\ref{eq:OmegaNumQM}) at hand, we can now study the structure of the phase diagram numerically.
Except for the analytically known averages of $M(z)$, $M(z)^2$ and $M'(z)^2$ given in Ref.~\cite{Schnetz:2005ih}, we evaluate all involved integrals numerically and obtain $\Omega_{MF,NJL/QM}(T,\mu;\Delta,\nu,\delta)$.
Since numerical minimizations typically determine parameters up to a total and/or relative error and since elliptic integrals depend on the elliptic modulus $\nu$ in a highly nonlinear fashion for $\nu\rightarrow 1$, it is numerically very helpful to express $\nu=1-16 \exp(-4\ln(2)/q_h)$ with $q_h\in [0,1]$ and then minimize in the parameters $\Delta$, $q_h$ and $\delta$. 

\noindent
We will now start discussing results for the NJL phase diagram in the chiral limit, i.e. for $m=0$. For the reasons discussed at the beginning of section~\ref{sec:QM}, we will fix $f_\pi=88\mathrm{MeV}$ to its value in the chiral limit and instead of adjusting $\langle\bar{\psi}\psi\rangle$, we choose a value for $M_q$. These choices in turn fix the model parameters through Eq.(\ref{eq:cbarcfpi}). On the left side of Fig.~\ref{fig1} we present the phase diagram restricting to homogeneous phases for a value of $M_q=300\mathrm{MeV}$.
For larger temperatures we see the second order phase transition line from the chirally broken to the restored phase, which turns into a first order phase transition line  at $(\mu_{cr}=269\mathrm{MeV},T_{cr}=74\mathrm{MeV})$ locating the chiral critical point.
The first order line then ends at  $(\mu=312\mathrm{MeV},T=0\mathrm{MeV})$, where the associated spinodal region spannes about $18\mathrm{MeV}$ in the quark chemical potential $\mu_q$.
Although the chiral condensate with $\langle\bar{\psi}\psi\rangle=-(193\mathrm{MeV})^3$ for this case is phenomenologically too small, we observe that the structure of the phase diagram is similar to that typically found in NJL models using a sharp three-momentum cutoff~\cite{Scavenius:2000qd}.

\noindent
Focusing on the region near/around the first order phase transition and the critical point, the right hand side of Fig.~\ref{fig1} shows the same lines as on the left, but now also including the domain where inhomogeneous phases are energetically preferred.
As discussed in Ref.~\cite{Nickel:2009ke} for the vicinity of the critical point, we observe that there is no longer a first order phase transition in the phase diagram, since it is replaced by an inhomogeneous ground state.
The transitions from the chirally broken to the inhomogeneous and from the inhomogeneous to the restored phase are both second order, where the first transition is characterized by the formation of (in the perpendicular direction) localized domain-wall solitons and the second by the melting of the condensate.

\begin{figure}
\includegraphics[width=7.5cm]{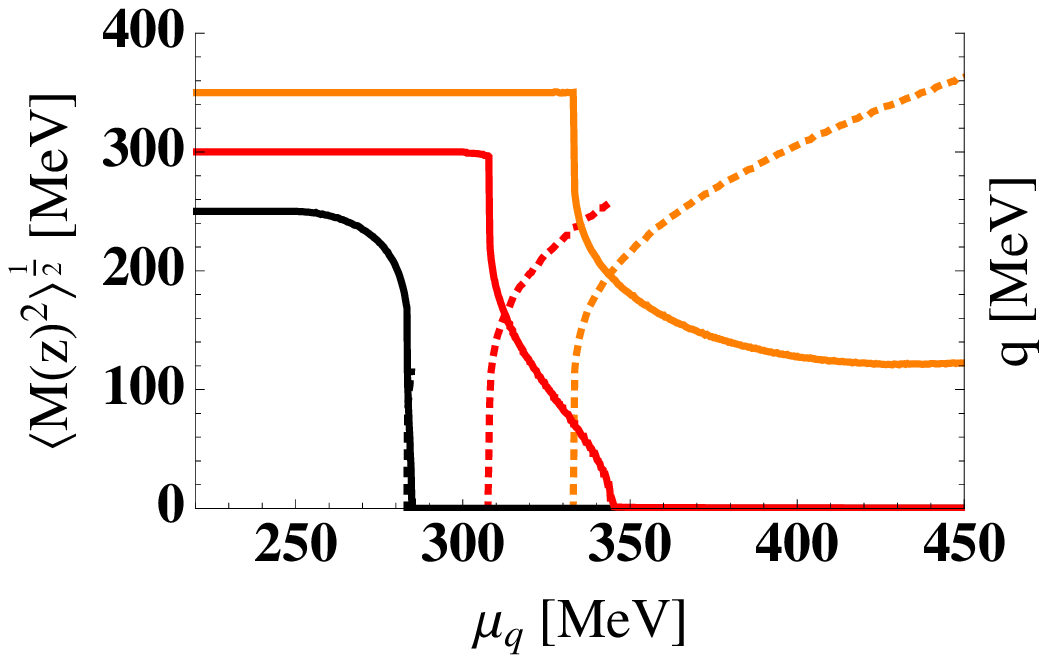}
\hspace{1cm}
\includegraphics[width=7.5cm]{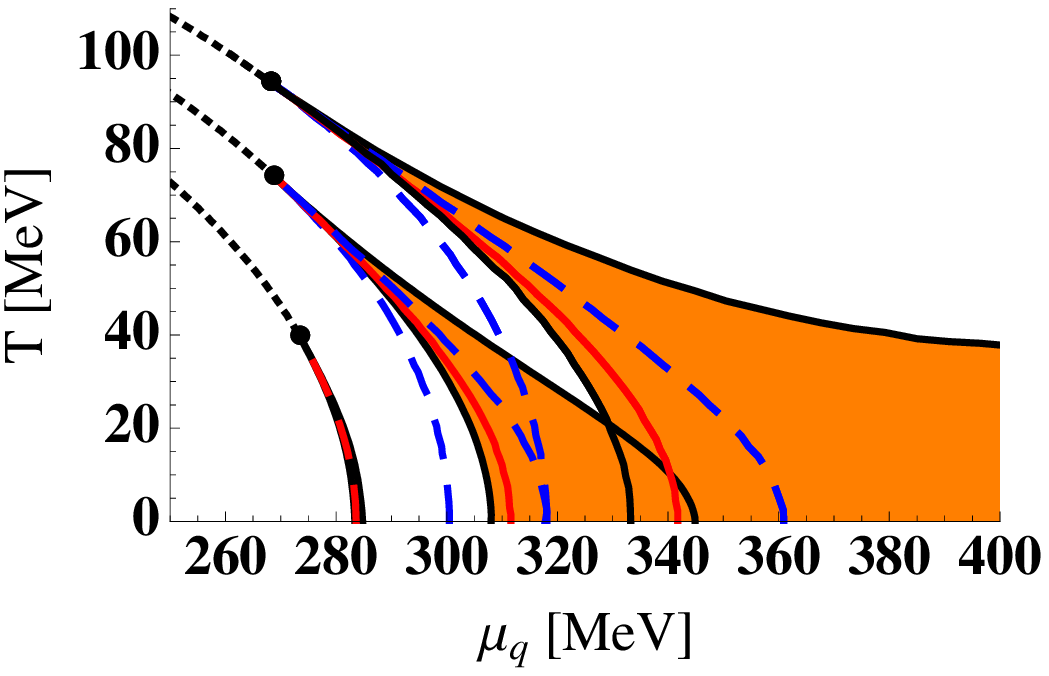}
\caption{
Left: Wave-vector $q$ (dashed lines) and average of constituent mass $\sqrt{\langle M(z)^2\rangle}$ (solid lines) at vanishing temperatures as function of quark chemical potential $\mu_q$ for $M_q=250\mathrm{MeV}$ (black lines), $M_q=300\mathrm{MeV}$ (red lines) and $M_q=350\mathrm{MeV}$ (orange lines).
Right: Same plot as on the right of Fig.~\ref{fig1}, now including results for $M_q=350\mathrm{MeV}$ (upper branch) and $M_q=250\mathrm{MeV}$ (lower branch).
}
\label{fig2}
\end{figure}

\noindent
The nature of the phase transitions is also apparent in the squared spatial average $\langle M(x)^2\rangle^\frac{1}{2}$ of the order parameter as well as the wave-vector of the one-dimensional modulation $q$. For vanishing temperatures these are depicted on the left of Fig.~\ref{fig2} for $M_q=250,300,350\mathrm{MeV}$.
We observe that $q$ continuously raises from $q=0$ at the transition from the chirally broken to the inhomogeneous phase, which is related to the formation of localized objects. On the other hand $\langle M(x)^2\rangle^\frac{1}{2}$ continuously goes to zero at the transition from the inhomogeneous to the chirally restored phase.
In the same plot we also see that at vanishing temperatures the constituent quark mass stays constant for $\mu_q\leq M_q$ (below any phase transition) and starts decreasing with the condensation of quarks.

\noindent
Whether or not quarks condense before reaching a phase transition has tremendous consequences on the structure of the phase diagram as shown on the right hand side of Fig.~\ref{fig2}.
Here we present the region of the first order phase transition for $M_q=250,300,350\mathrm{MeV}$. Qualitatively the picture for those three cases is similar, but quantitatively there are significant differences.
Since for $M_q=250\mathrm{MeV}$ the quarks already form a Fermi surface before reaching the phase transition, the order parameter is already diminished at the phase transition. As a result the phase transition is very weak and the spinodal region/domain of inhomogeneous phases strongly reduced.
For $M_q=350\mathrm{MeV}$ on the other side the interaction is very strong and there is no condensation before reaching the onset of inhomogeneous phase (or the first order phase transition when limiting to homogeneous phases).
The domain of inhomogeneous phases is very large here and it is conceivable that its persistence beyond $\mu_q=400\mathrm{MeV}$ is a regularization artefact.
We also point out that although the value of constituent quark mass $M_q$ is comparable to that in NJL models using a three-momentum cutoff, the 
chiral condensate with $\langle\bar{\psi}\psi\rangle=-(186\mathrm{MeV})^3$ at $M_q=350\mathrm{MeV}$ decreases with increasing $M_q$ here.

\begin{figure}
\includegraphics[width=7.5cm]{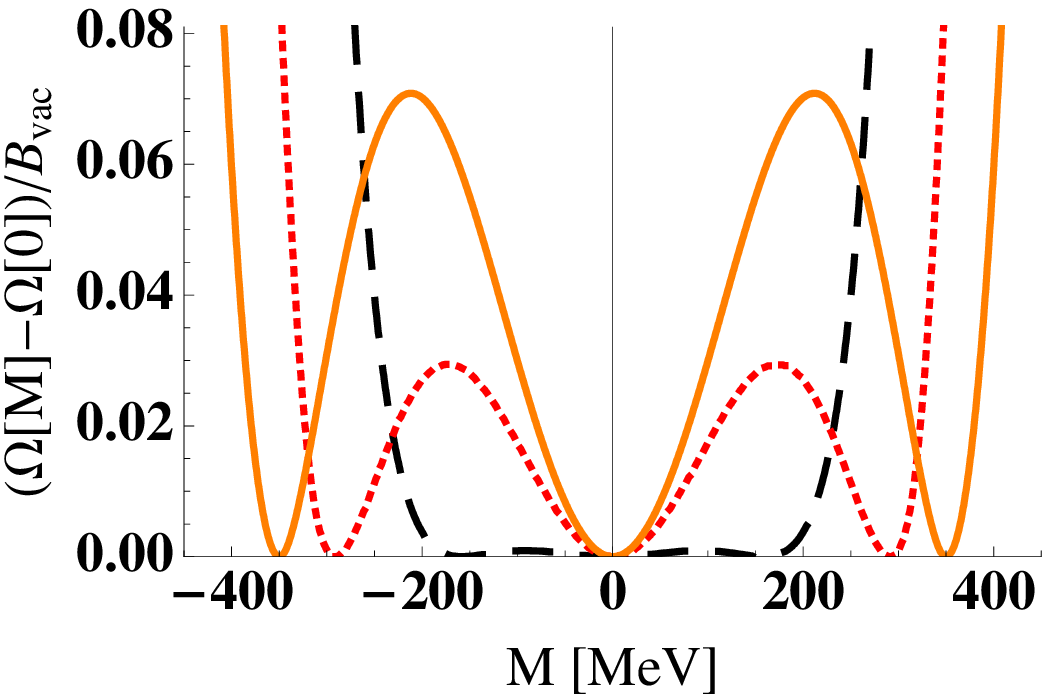}
\hspace{1cm}
\includegraphics[width=7.5cm]{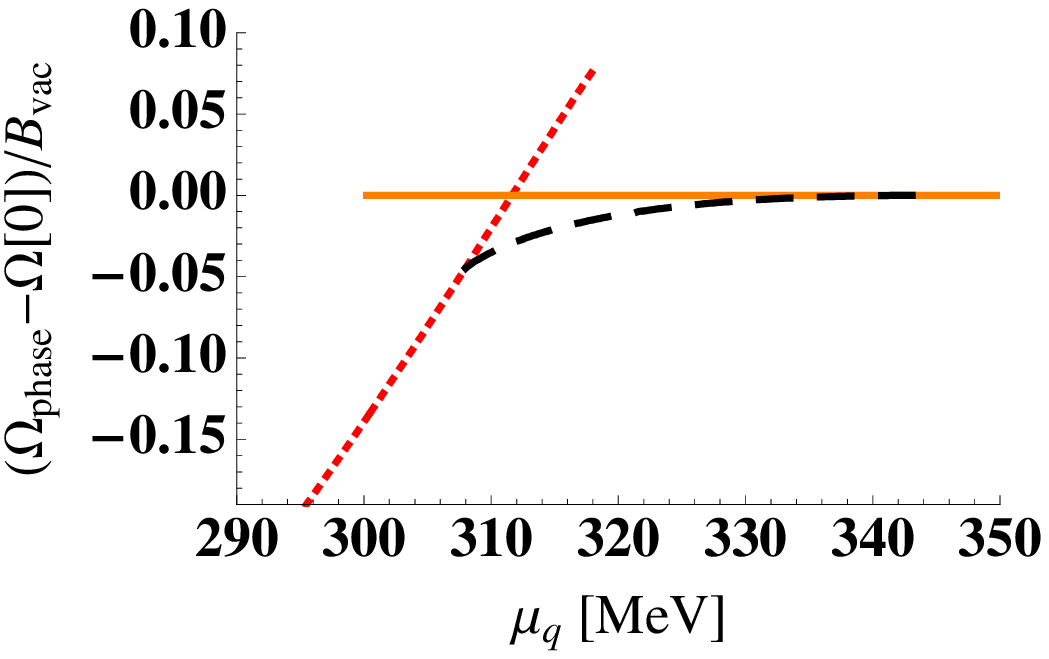}
\caption{
Left: Difference of thermodynamic potential $\Omega(T,\mu;M)$ in the chiral limit for the homogeneous phase with constituent mass $M$ and its value for $M=0$ in units of the bag constant $B_{vac}$.
The black (long-dashed) line corresponds to $M_q=250\mathrm{MeV}$, the red (short-dashed) to $M_q=300\mathrm{MeV}$ and the orange (solid) to $M_q=350\mathrm{MeV}$.
All for vanishing temperatures and $\mu_q$ at the first order phase transition (when limiting to homogeneous phases).
Right: Difference of thermodynamic potential $\Omega_{phase}$ in the chiral limit for various phases and its value for $M=0$ for $M_q=300\mathrm{MeV}$ in units of the bag constant $B_{vac}$. The black (long-dashed) line corresponds to the inhomogeneous, the red (short-dashed) to the chirally broken and the orange (solid) to the chirally restored phase.
}
\label{fig3}
\end{figure}

\noindent
The results indicate that the size of the domain for inhomogeneous ground-states, similar to the spinodal region, is related to the strength of the first order phase transition. In order to give a better picture for the latter, the thermodynamic potentials as effective actions in the order parameter are shown on the left of Fig.~\ref{fig3} at vanishing temperatures and the respective first order phase transition points.
We find that the local maxima between $M_q=250\mathrm{MeV}$ and $M_q=350\mathrm{MeV}$ in the potential even when measured in the respective bag constants $B_{vac}=\Omega(0,0;0)-\Omega(0,0;M_q)$ differs by more than an order of magnitude, therefore underlining the quantitative difference between the two cases.
For completeness, the value of the bag constant for $M_q=250,300,350\mathrm{MeV}$ is $B_{vac}=40,62,91\mathrm{MeV/fm^3}$, respectively.

\noindent
Coming back to $M_q=300\mathrm{MeV}$ we also like to give an impression of the thermodynamic potential at vanishing temperatures using the right hand side of Fig.~\ref{fig3}.
Here we show the value of the thermodynamic potential for the energetically most preferred homogeneous phase as well as inhomogeneous phase. Plotted are only cases that form a local minimum of the thermodynamic potential as an effective action, such that e.g. the spinodal region can be deduced from the plot.
In the case of several local minima at a given $\mu_q$ the global minimum is energetically preferred and we see that if existent, the inhomogeneous phase forms the ground state.
Also the order of the phase transitions can be deduced from the plot.

\begin{figure}
\includegraphics[width=7.5cm]{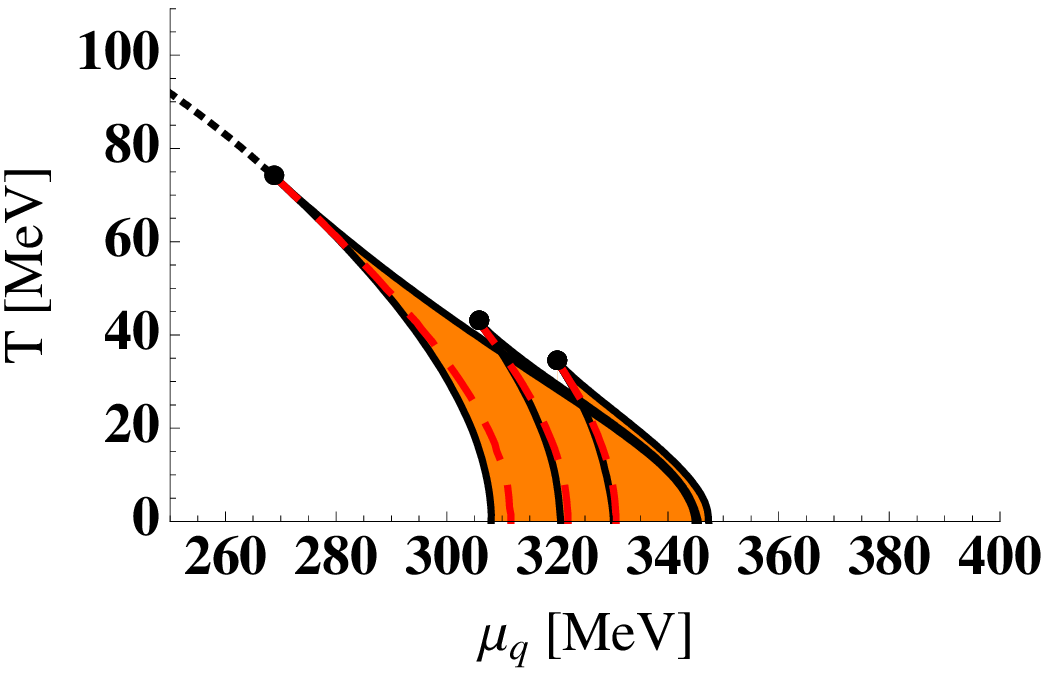}
\hspace{1cm}
\includegraphics[width=7.5cm]{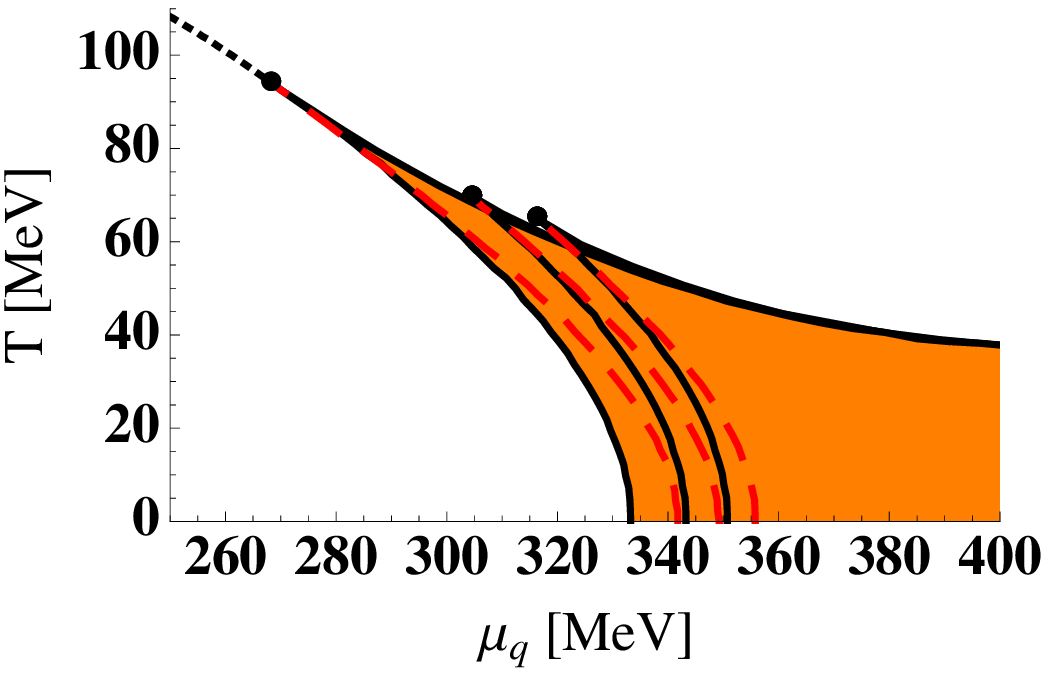}
\caption{
Left: Same plot as on the right of Fig.~\ref{fig1}, now including the domain of inhomogeneous phases for $m=5\mathrm{MeV}$ and $m=10\mathrm{MeV}$. Branches with critical points at smaller temperature $T$ and larger quark chemical potential $\mu_q$ correspond to larger current quark masses $m$.
Right: Same plot as on the left for $M_q=350\mathrm{MeV}$.
}
\label{fig4}
\end{figure}

\noindent
As a next point we want to address the case of finite current quark masses. For this purpose we again fix the parameters $G_s$, $\Lambda$ by choosing $f_\pi$, $M_q$ in the chiral limit and then turn on a finite current quark mass $m$.
In Fig.~\ref{fig4} we present our results for the relevant part of the phase diagram choosing $m=0, 5, 10\mathrm{MeV}$.
Aside from the fact that the second order phase transition is turned into a cross-over when leaving the chiral limit, we see that the qualitative picture stays the same: Again the first order phase transition line is replaced by two second order phase transition lines which enclose a domain where inhomogeneous phases are energetically preferred.
All lines meet at the critical point, which can be calculated on the bases of homogeneous phases only and which is known to shift towards smaller temperatures and larger quark chemical potentials when increasing $m$.

\begin{figure}
\includegraphics[width=7.5cm]{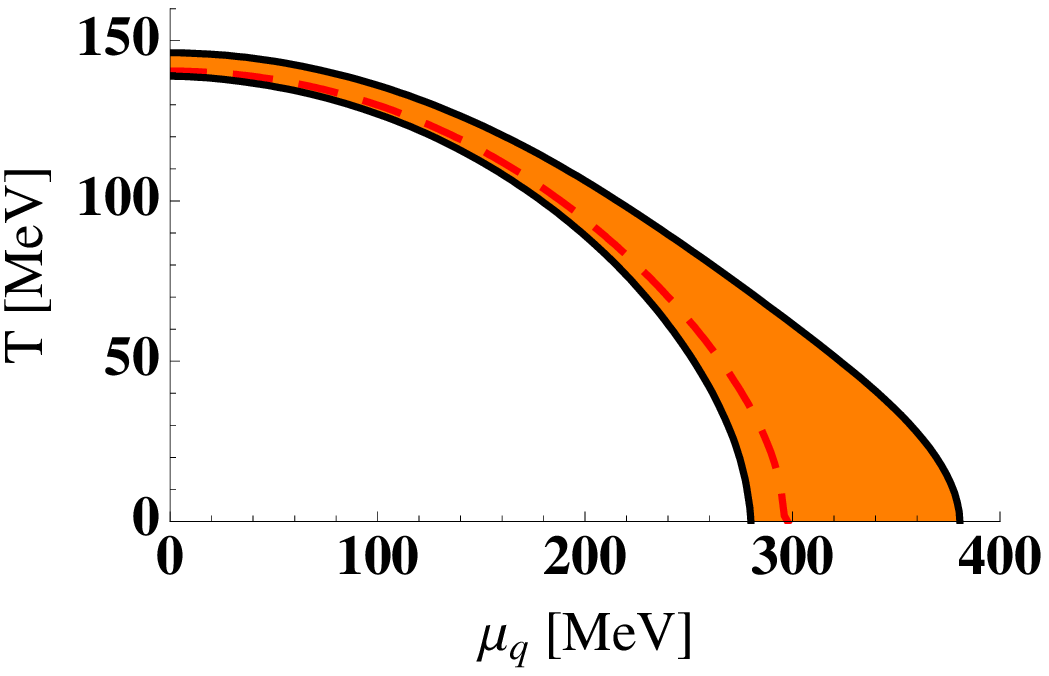}
\hspace{1cm}
\includegraphics[width=7.5cm]{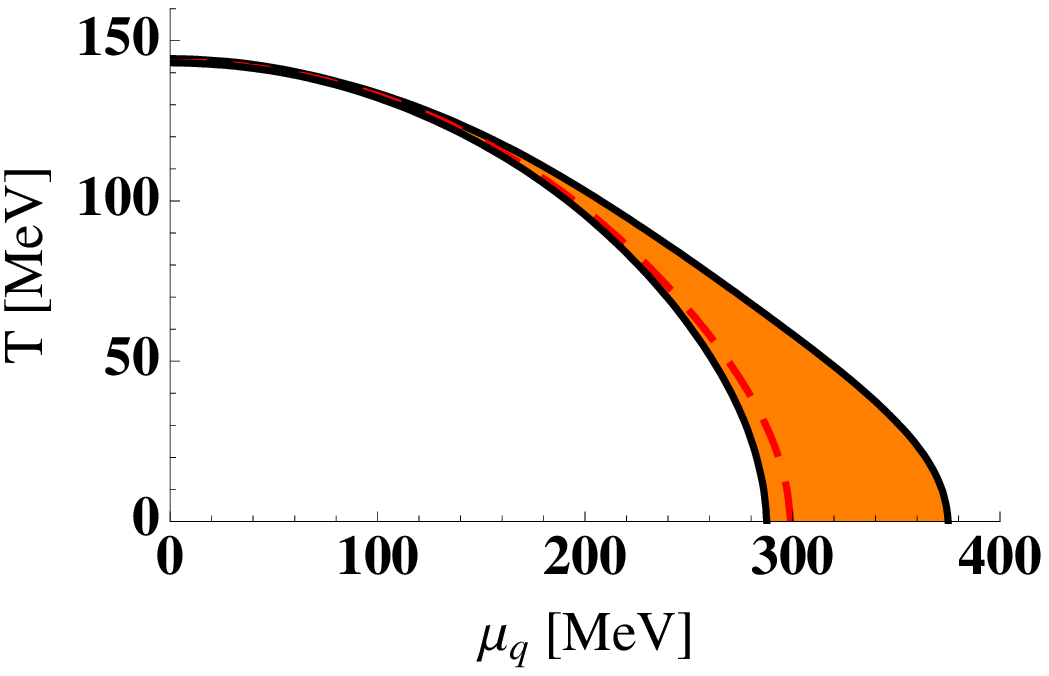}
\caption{
Left: Phase diagram for the QM model in the chiral limit with the red (dashed) line indicating the first order phase transition when limiting to homogeneous phases only. Inhomogeneous phases are preferred in the orange (shaded) domain enclosed by black (solid) lines.
Right: Same plot as on the left, now for $m_\pi=69\mathrm{MeV}$.
}
\label{fig5}
\end{figure}

\begin{figure}
\includegraphics[width=7.5cm]{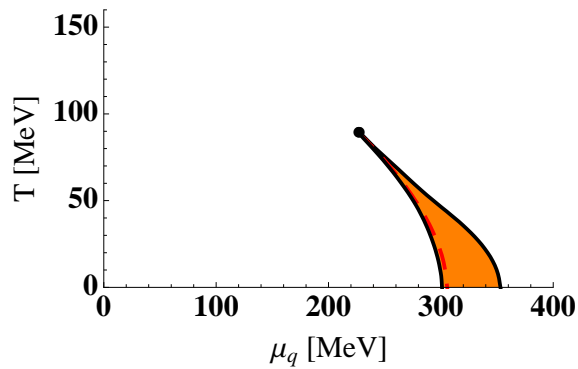}
\caption{
Same plot as on the left of Fig.~\ref{fig5}, now for $m_\pi=138\mathrm{MeV}$ and with the bullet showing the critical point.
}
\label{fig6}
\end{figure}

\noindent
At the end of this section we sketch some results for the QM model and due to the similarity to the NJL model we will restrict to the structure of the phase diagram only.
Following Refs.~\cite{Scavenius:2000qd,Schaefer:2006ds} we choose the model parameters by $f_\pi=93\mathrm{MeV}$, $M_q=300\mathrm{MeV}$, $m_\sigma=600\mathrm{MeV}$ and vary the pion mass $m_\pi$.
In Figs.~\ref{fig5},\ref{fig6} we show the phase diagrams including the domain for inhomogeneous phases with $m_\pi=0, 69,1 38\mathrm{MeV}$.
The model in the applied approximation does not have a critical point in the chiral limit and as expected after the discussion of the model in section~\ref{sec:QM} and the numerical results for the NJL model, the first order phase transition from the chirally broken to the (partially) restored phase is replaced by two second order phase transitions at the boundary of a inhomogeneous domain.
For half the physical pion mass $m_\pi=69\mathrm{MeV}$ we find the same scenario, whereas we find a critical point in the phase diagram at the physical value of the pion mass $m_\pi=138\mathrm{MeV}$.
Although the critical point is in principle hidden by inhomogeneous phases as discussed at the end of section~\ref{sec:QM}, the Lifshitz point at the cusp of the two second order phase transition lines and the critical point appear to be so close that we cannot distinguish those numerically.

\section{discussion}
\label{sec:discussion}
\noindent
In this paper we have shown how inhomogeneous phases in the NJL  and the QM model with lower dimensional modulations can be investigated and especially how this analysis is related that in the (chiral) Gross-Neveu model.
As suggested in Ref.~\cite{Nickel:2009ke} we have confirmed that the first order phase transition line in the NJL phase diagram of homogeneous phases is completely replaced by two second order phase transition lines that border an inhomogeneous phase.
The intersection of these lines defines a Lifshitz point which coincides with the critical point also for finite current quark masses.
For the QM model in the applied approximation this picture is somewhat modified as the critical point (in the case of homogeneous phases) is slightly hidden by an inhomogeneous phase.
Although the qualitative behavior is very general, the specific structure of the phase diagram and in particular the importance of inhomogeneous phases is strongly depending on the model parameters.
We have established that there is a close connection between the significance of inhomogeneous phases and the strength of the first order phase transition when limiting to homogeneous phases, in particular the size of its spinodal region.
In addition also the relevance of pseudo-scalar condensates has been addressed.

\noindent
This work is a further step in the investigation of inhomogeneous phases within NJL-type models. In the following various related problems could be addressed:
From a phenomenological point of view it would be interesting to study the interplay and competition with color-superconducting phases, being expected in the same domain of the phase diagram~\cite{reviews,Alford:2007xm}.
Also the calculation of higher-dimensional modulations should be in range of numerical approaches, especially since the magnitude of the order parameter and the size of the Brillouin zone can be large and a numerical determination of the density of states therefore be possible~\cite{Nickel:2008ng}.
More towards an extension of the considered models, a GL analyses including e.g. vector-vector and 't Hooft interaction could be performed to check whether and how the relation of Lifshitz and chiral critical point can be modified.
Also inhomogeneous phases in the Polyakov-NJL model can be studied easily since the density of states for quasi-particles is not altered by the inclusion of the Polyakov loop. As a consequence the technical issues associated with inhomogeneous phases are not furtherly complicated here.
From a technical point of view e.g. the zero point energies in the QM model can be easily taken into account since the asymptotic behavior of the spectral density is known analytically. This would then allow for a proper renormalization of the model.
In general it would of course also be interesting to study these phases beyond mean-field approximation, especially since fluctuations play a key role at a critical point.

\acknowledgments
\noindent
We thank G. Basar, M. Buballa, S. Carignano, G. Dunne, K. Rajagopal and M. Stephanov for helpful comments and discussions.
This work was supported in part by funds provided by the U.S. Department of Energy (D.O.E.) under cooperative research agreement DE-FG0205ER41360 and by the German Research Foundation (DFG) under grant number Ni 1191/1-1.



\begin{thebibliography}{99}

\bibitem{reviews}
  K.~Rajagopal and F.~Wilczek,
  arXiv:hep-ph/0011333;
  T.~Sch\"afer,
  arXiv:hep-ph/0304281;
  D.~H.~Rischke,
  Prog.\ Part.\ Nucl.\ Phys.\  {\bf 52}, 197 (2004)
  [arXiv:nucl-th/0305030];
  M.~Buballa,
  Phys.\ Rept.\  {\bf 407}, 205 (2005)
  [arXiv:hep-ph/0402234].

\bibitem{Alford:2007xm}
  M.~G.~Alford, A.~Schmitt, K.~Rajagopal and T.~Schafer,
  Rev.\ Mod.\ Phys.\  {\bf 80}, 1455 (2008)
  [arXiv:0709.4635 [hep-ph]].

\bibitem{Stephanov:2007fk}
  M.~A.~Stephanov,
  PoS {\bf LAT2006}, 024 (2006)
  [arXiv:hep-lat/0701002].

\bibitem{Fukushima:2003fw}
  K.~Fukushima,
  Phys.\ Lett.\  B {\bf 591}, 277 (2004)
  [arXiv:hep-ph/0310121].

\bibitem{Ratti:2005jh}
  C.~Ratti, M.~A.~Thaler and W.~Weise,
  Phys.\ Rev.\  D {\bf 73}, 014019 (2006)
  [arXiv:hep-ph/0506234].

\bibitem{Nambu:1961tp}
  Y.~Nambu and G.~Jona-Lasinio,
  Phys.\ Rev.\  {\bf 122}, 345 (1961).

\bibitem{GellMann:1960np}
  M.~Gell-Mann and M.~Levy,
  Nuovo Cim.\  {\bf 16}, 705 (1960).

\bibitem{Scavenius:2000qd}
  O.~Scavenius, A.~Mocsy, I.~N.~Mishustin and D.~H.~Rischke,
  Phys.\ Rev.\  C {\bf 64}, 045202 (2001)
  [arXiv:nucl-th/0007030].

\bibitem{Deryagin:1992rw}
  D.~V.~Deryagin, D.~Y.~Grigoriev and V.~A.~Rubakov,
  Int.\ J.\ Mod.\ Phys.\  A {\bf 7}, 659 (1992).

\bibitem{Shuster:1999tn}
  E.~Shuster and D.~T.~Son,
  Nucl.\ Phys.\  B {\bf 573}, 434 (2000)
  [arXiv:hep-ph/9905448].

\bibitem{Park:1999bz}
  B.~Y.~Park, M.~Rho, A.~Wirzba and I.~Zahed,
  Phys.\ Rev.\  D {\bf 62}, 034015 (2000)
  [arXiv:hep-ph/9910347].

\bibitem{Rozali:2007rx}
  M.~Rozali, H.~H.~Shieh, M.~Van Raamsdonk and J.~Wu,
  JHEP {\bf 0801}, 053 (2008)
  [arXiv:0708.1322 [hep-th]].

\bibitem{McLerran:2007qj}
  L.~McLerran and R.~D.~Pisarski,
  Nucl.\ Phys.\  A {\bf 796}, 83 (2007)
  [arXiv:0706.2191 [hep-ph]].

\bibitem{Rapp:2000zd}
  R.~Rapp, E.~V.~Shuryak and I.~Zahed,
  Phys.\ Rev.\  D {\bf 63}, 034008 (2001)
  [arXiv:hep-ph/0008207].

\bibitem{Sadzikowski:2000ap}
  M.~Sadzikowski and W.~Broniowski,
  Phys.\ Lett.\  B {\bf 488}, 63 (2000)
  [arXiv:hep-ph/0003282].

\bibitem{Nakano:2004cd}
  E.~Nakano and T.~Tatsumi,
  Phys.\ Rev.\  D {\bf 71}, 114006 (2005)
  [arXiv:hep-ph/0411350].

\bibitem{Nickel:2009ke}
  D.~Nickel,
  arXiv:0902.1778 [hep-ph].

\bibitem{Schnetz:2004vr}
  O.~Schnetz, M.~Thies and K.~Urlichs,
  Annals Phys.\  {\bf 314}, 425 (2004)
  [arXiv:hep-th/0402014].

\bibitem{Schnetz:2005ih}
  O.~Schnetz, M.~Thies and K.~Urlichs,
  Annals Phys.\  {\bf 321}, 2604 (2006)
  [arXiv:hep-th/0511206].

\bibitem{Thies:2006ti}
  M.~Thies,
  J.\ Phys.\ A  {\bf 39}, 12707 (2006)
  [arXiv:hep-th/0601049].

\bibitem{Correa:2009xa}
  F.~Correa, G.~V.~Dunne and M.~S.~Plyushchay,
  arXiv:0904.2768 [hep-th].

\bibitem{Basar:2009fg}
  G.~Basar, G.~V.~Dunne and M.~Thies,
  arXiv:0903.1868 [hep-th].

\bibitem{Itzykson:1980rh}
  C.~Itzykson and J.~B.~Zuber,
{\it  New York, Usa: Mcgraw-hill (1980) 705 P.(International Series In Pure and Applied Physics)}

\bibitem{Basar:2008im}
  G.~Basar and G.~V.~Dunne,
  Phys.\ Rev.\ Lett.\  {\bf 100}, 200404 (2008)
  [arXiv:0803.1501 [hep-th]];
   G.~Basar and G.~V.~Dunne,
  Phys.\ Rev.\  D {\bf 78}, 065022 (2008)
  [arXiv:0806.2659 [hep-th]].

\bibitem{Klevansky:1992qe}
  S.~P.~Klevansky,
  Rev.\ Mod.\ Phys.\  {\bf 64}, 649 (1992).

\bibitem{Boehmer:2007ea}
  C.~Boehmer, M.~Thies and K.~Urlichs,
  Phys.\ Rev.\  D {\bf 75}, 105017 (2007)
  [arXiv:hep-th/0702201].

\bibitem{LO64}   
  A.~I.~Larkin and Yu.~N.~Ovchinnikov, 
  Zh.\ Eksp.\ Teor.\ Fiz.\ {\bf 47}, 1136 (1964); 
  Sov.\ Phys.\ JETP {\bf 20} 762 (1965).     

\bibitem{Bowers:2002xr}
  J.~A.~Bowers and K.~Rajagopal,
  Phys.\ Rev.\  D {\bf 66}, 065002 (2002)
  [arXiv:hep-ph/0204079].

\bibitem{Schaefer:2006ds}
  B.~J.~Schaefer and J.~Wambach,
  Phys.\ Rev.\  D {\bf 75}, 085015 (2007)
  [arXiv:hep-ph/0603256].

\bibitem{Nickel:2008ng}
  D.~Nickel and M.~Buballa,
  Phys.\ Rev.\  D {\bf 79}, 054009 (2009)
  [arXiv:0811.2400 [hep-ph]].



\end{thebibliography}
\end{document}